# TREND VALIDATION OF A MUSCULOSKELETAL MODEL WITH A WORKSTATION DESIGN PARAMETER

Charles Pontonnier[1,2], Mark de Zee[1], Afshin Samani[1], Georges Dumont[2,3] and Pascal Madeleine[1]

[1] Center for Sensory-Motor Interaction (SMI), Department of Health Science and Technology, DK-9220 Aalborg, Denmark
[2] VR4I project team, IRISA, Campus de Beaulieu, F-35042 Rennes Cédex, France
[3] ENS Cachan Antenne de Bretagne, Campus de Ker Lann, F-35170 Bruz, France
Email: cpontonn@hst.aau.dk

## INTRODUCTION
The aim of this article is to present the application of a trend validation to validate a simulation model. The workstation parameter used to define the trend is the table height of simulated meat cutting tasks (well known to be related to MSD [1]).

## METHODS
In this article, we investigated the effect of the table height on muscle activation levels during a meat cutting task. Preliminary results are provided from a single pilot recording (Age: 39 yrs; Body Mass: 72 kg; Height: 1.86 m). The subject was standing in front of a table and performing 10 cutting task cycles in the designed directions at three different table heights. The first one is the recommended table height for a light work [2] that corresponds to elbow height for a neutral position of the arm and was used as a reference. The two other heights were 10 and 20 cm below the reference height. The investigation aimed at verifying the musculoskeletal simulation by comparing simulated activation trends and the recorded EMG amplitude when the table height is decreasing.

The data collected during the motion consists of movement data (Visualeyez II™ system set up with two VZ4000 trackers, Phoenix Technologies Inc., BC, Canada) sampled at 60 Hz. Twelve active markers were used to collect trunk, right arm and knife motion with respect to anatomical landmarks. Trajectories were low-pass filtered (Butterworth 2nd order, $F_{cut\text{-}off}$ 5 Hz). To register the applied external forces, we used an instrumented knife that allows the recording of the cutting force in 3D. The knife is based on a 3D force sensor (FS6, AMTI, Watertown, MA, USA). Force signals were low-pass filtered (10.5 Hz) and amplified 2000 times. The signals were A/D converted and sampled at 60 Hz (12 bits A/D converter, Nidaq 6024, National Instruments, Austin, TX, USA) and recorded through a custom made program (Mr Kick, Aalborg University, Aalborg) in LabView 8.2 (National Instruments, Austin, TX,USA), which also provided feedback to the experimenter.

Both motion capture and external force data were used to drive a musculoskeletal model of the trunk and the arm (52 degrees of freedom and 146 muscles) in the AnyBody Software [3]. For each table height, we only computed one frame corresponding to the mean half time of the task with a constant force of 30N (applied in the recorded direction during the real task) to obtain simulated muscle activations in the shoulder/neck area. Figure 1 shows the real experimentation task and the simulated one.

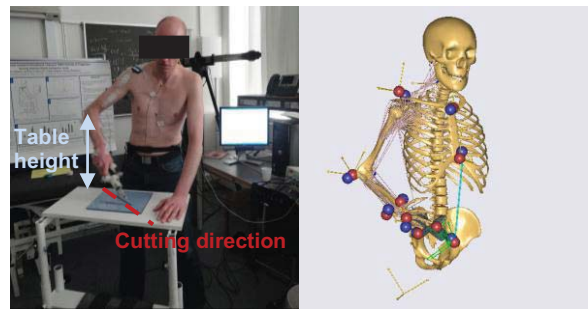

**Fig 1**: Real task experimental protocol and simulated task in the AnyBody software.

To verify if the trend followed by simulated muscle activations is matching with actual exerted force, we recorded electromyographic activation of several superficial muscles. A 64-channel matrix was used to collect the muscle activity of the upper-middle trapezius. Moreover, four bipolar channels were used to collect EMG from the Deltoideus Medialis, Deltoideus Anterior, Biceps Brachii and Triceps Long head with bipolar surface electrodes. Bipolar electrodes were placed with respect to anatomical landmarks. The EMG signals were amplified 2000 times (64-channel surface EMG amplifier, SEA64, LISiN-OT Bioelectronica, Torino, Italy) and sampled at 2048

Hz (National Instrument, 12 bits acquisition board, Austin, USA). Signals were low pass filtered (4 Hz), and rectified – using the envelope as final value.

**RESULTS AND DISCUSSION**

Table 1 presents the trend followed by the measured activations (R) and the simulated activations (S) for three table heights. Recorded activations are the mean activations extracted from 10 cutting cycles and the simulated activations are obtained for a unique frame driven at the same height from motion capture data, and for a constant input force of 30N in the recorded cutting direction. 0cm height is used as a reference for the activations. The activations were reported as the percentage of its difference from the corresponding reference.

| Height | 0cm | -10cm | -20cm |
|---|---|---|---|
| Biceps R | 0% | -24,4% | -36,5% |
| Biceps S | NA | NA | NA |
| Triceps R | 0% | +7,3% | -3,7% |
| Triceps S | 0% | +18,84% | -13,0% |
| Delt Ant R | 0% | -9,2% | -30,9% |
| Delt Ant S | 0% | -18,0% | -36,0% |
| Delt Med R | 0% | -21,5% | -35,8% |
| Delt Med S | 0% | -18,0% | -36,0% |
| Trap Acr-C7 R | 0% | +9,8% | +10,6% |
| Trap Acr-C7 S | 0% | +3,45% | +17,2% |

**Table 1**: Comparison between the trends followed by recorded activations (R) and simulated activations (S) for the three table heights. The trapezius Acr-C7 is the activation of the muscle in the Acromion-C7 direction, computed as the mean of the 4 bipolar channels available in this direction.

First, absolute EMG envelopes showed that the decreased table height led to a global decrease of the mean recorded muscle activations for a ~30N output cutting force. For the simulated task, we can see that biceps is not activated (due to the fact that it is antagonistic to the output force) and cannot be used to investigate the trend. The trend is the same as the recorded one for Deltoidus Medialis and the Deltoidus Anterior (activations decrease in a similar range for both R and S). For the triceps, we can see that the recorded and the simulated activation levels do not follow the global trend, and shows that the -10cm height leads to a maximal activation compared to the two other ones in both cases (R and S). The simulation seems to overestimate the variations of the triceps activation during the task. It could be due to the fact that the biceps is not activated (activating the biceps will decrease the relative variation of the triceps activation by adding an offset on the triceps activation [4]).

The trend followed by the trapezius activation is also difficult to investigate, because we compare a recorded spatial activation to several sub-muscles activations (representing the trapezius in the AnyBody model). Here we compare the activation obtained on the action line between the acromion and the 7$^{th}$ cervical vertebra that corresponds to the location of the 4$^{th}$ line of the 64-channel matrix. We can see that the both R and S activations are following the same trend, but the variations are not similar in terms of range.

These results show that Deltoidus is the most reliable muscle to investigate for this workstation parameter. Both R and S for anterior and medial part are following a similar trend. This is why in further analysis we will use these activations to investigate the effect of the table height.

Finally, simulated and real activations suggest that the table height has to be set slightly below the elbow height for a neutral position of the arm to decrease the muscle activation levels.

Further analysis and investigations are needed, especially with respect to the kinematics during the task. Kinematics parameters such as range of motion are highly related to the muscle activation levels. Finally, these results have also to be validated on a relevant statistical sample of subjects. The 64-channel matrix can at last be used to investigate muscle fatigue and pain phenomenon [5].

**CONCLUSIONS**

This short article based on a pilot recording presents a trend comparison between recorded and simulated activations with respect to a workstation design parameter (the table height) during meat cutting tasks. Results show that the trend followed by the simulated activations is the same as the one followed by the recorded data. The most reliable activations are the Deltoidus ones. Antagonistic muscle activations are difficult to investigate with a classic inverse dynamics method; this is why in further investigation we propose to simulate the task with a hybrid (inverse-dynamics assisted by EMG) method.